\documentclass[reprint,superscriptaddress,amsmath,amssymb,aps,pra,floatfix]{revtex4-1}

\usepackage[dvips]{graphicx}
\usepackage[usenames]{color}
\usepackage{dcolumn}
\usepackage{bm}
\usepackage{epstopdf}
\usepackage{multirow}
\usepackage[normalem]{ulem}
\usepackage{hyperref}
\hypersetup{colorlinks   = true,linkcolor=blue,citecolor = blue,urlcolor=blue}

\begin{document}

\title{Superconductivity behavior in epitaxial TiN films points to surface magnetic disorder}
\author{N.A.~Saveskul}
\affiliation{Moscow State University of Education, 29 Malaya Pirogovskaya St, Moscow, 119435, Russia}
\author{N.A.~Titova}
\affiliation{Moscow State University of Education, 29 Malaya Pirogovskaya St, Moscow, 119435, Russia}
\author{E.M.~Baeva}
\affiliation{National Research University Higher School of Economics, 20 Myasnitskaya St, Moscow, 101000, Russia}
\affiliation{Moscow State University of Education, 29 Malaya Pirogovskaya St, Moscow, 119435, Russia}
\author{A.V.~Semenov}
\affiliation{Moscow State University of Education, 29 Malaya Pirogovskaya St, Moscow, 119435, Russia}
\author{A.V.~Lubenchenko}
\affiliation{National Research University MPEI, Krasnokazarmennaya St., 14, Moscow, 111250, Russia}
\author{S.~Saha}
\affiliation{School of Electrical \& Computer Engineering and Birck Nanotechnology Center, Purdue University, 1205 West State Street, West Lafayette, Indiana 47907-2057, USA}
\affiliation{Purdue Quantum Science and Engineering Institute, Purdue University, West Lafayette, Indiana 47907, USA}
\author{H.~Reddy}
\affiliation{School of Electrical \& Computer Engineering and Birck Nanotechnology Center, Purdue University, 1205 West State Street, West Lafayette, Indiana 47907-2057, USA}
\affiliation{Purdue Quantum Science and Engineering Institute, Purdue University, West Lafayette, Indiana 47907, USA}
\author{S.~Bogdanov}
\affiliation{School of Electrical \& Computer Engineering and Birck Nanotechnology Center, Purdue University, 1205 West State Street, West Lafayette, Indiana 47907-2057, USA}
\affiliation{Purdue Quantum Science and Engineering Institute, Purdue University, West Lafayette, Indiana 47907, USA}
\author{E.E.~Marinero}
\affiliation{School of Electrical \& Computer Engineering and Birck Nanotechnology Center, Purdue University, 1205 West State Street, West Lafayette, Indiana 47907-2057, USA}
\affiliation{School of Materials Engineeing, Purdue University, 1205 West State Street, West Lafayette, Indiana 47907-2057, USA}
\author{V.M.~Shalaev}
\affiliation{School of Electrical \& Computer Engineering and Birck Nanotechnology Center, Purdue University, 1205 West State Street, West Lafayette, Indiana 47907-2057, USA}
\affiliation{Purdue Quantum Science and Engineering Institute, Purdue University, West Lafayette, Indiana 47907, USA}
\author{A.~Boltasseva}
\affiliation{School of Electrical \& Computer Engineering and Birck Nanotechnology Center, Purdue University, 1205 West State Street, West Lafayette, Indiana 47907-2057, USA}
\affiliation{Purdue Quantum Science and Engineering Institute, Purdue University, West Lafayette, Indiana 47907, USA}
\author{V.S.~Khrapai}
\affiliation{National Research University Higher School of Economics, 20 Myasnitskaya St, Moscow, 101000, Russia}
\affiliation{Moscow State University of Education, 29 Malaya Pirogovskaya St, Moscow, 119435, Russia}
\affiliation{Insitute of Solid State Physics, 2 Ak. Osipyana St., Chernogolovka, 142432, Russia}
\author{A.I.~Kardakova}
\affiliation{National Research University Higher School of Economics, 20 Myasnitskaya St, Moscow, 101000, Russia}
\affiliation{Moscow State University of Education, 29 Malaya Pirogovskaya St, Moscow, 119435, Russia}
\author{G.N.~Goltsman}
\affiliation{National Research University Higher School of Economics, 20 Myasnitskaya St, Moscow, 101000, Russia}
\affiliation{Moscow State University of Education, 29 Malaya Pirogovskaya St, Moscow, 119435, Russia}




\begin{abstract}
We analyze the evolution of the normal and superconducting electronic properties in epitaxial TiN films, characterized by high Ioffe-Regel parameter values, as a function of the film thickness. As the film thickness decreases, we observe an increase of the residual resistivity, which becomes dominated by diffusive surface scattering for $d\leq20$\,nm. At the same time, a substantial thickness-dependent reduction of the superconducting critical temperature is observed compared to the bulk TiN value. In such a high quality material films, this effect can be explained by a weak magnetic disorder residing in the surface layer with a characteristic magnetic defect density of $\sim10^{12}\,\mathrm{cm}^{-2}$. Our results suggest that surface magnetic disorder is generally present in oxidized TiN films.

\end{abstract}

\maketitle

\section{\label{introduction}Introduction}

Thin metallic films are exploited in numerous optical applications from nanophotonics and telecommunications at a room temperature~\citep{Maiern2007, Catellani2017} to superconducting electronic devices at cryogenic temperatures~\citep{Chang2015, Yan2018}. Critical for optical and electronic applications, improving the film quality is a multifaceted problem that includes dealing with various disorder types that have different impacts on the electronic properties at ambient conditions and on the superconducting state. A classical example is the effect of paramagnetic impurities in metals, where a  minute concentration of impurities can become detrimental at low temperature ($T$), resulting in a Kondo effect~\citep{MH71}, the suppression of the superconducting gap~\cite{AG61} and a drastic enhancement of the inelastic scattering~\cite{Pierre2003}. In thin films, a more important effect is produced by magnetic disorder formed spontaneously within oxidized native surface layers, which manifests in enhanced dephasing~\cite{Vranken1988,Pierre2002}, Cooper-pair breaking~\citep{Rogachev06, Proslier08} and magnetic flux noise~\cite{Anton2013,Kumar2016}.

Titanium nitride (TiN) thin films exhibit good chemical stability down to nanometer thickness~\citep{Chawla2013} and are used in the fabrication of superconducting devices for photon detection~\citep{Leduc10} and for quantum information processing~\citep{Ohya14,Makise15,Tang16,Foxen18}. Low  dielectric losses at microwave frequencies observed in TiN films are associated with a relatively small surface density of two-level systems defects that contribute to decoherence of the resonators and qubits~\citep{Vissers10, Sandberg2012, Chang13, Calusine18}. In spite of a possible relation between the two-level systems and the magnetic disorder~\cite{deGraaf2018}, the impact of the latter in TiN films is much less understood. Although the experiments do not exclude an unknown time-reversal symmetry breaking mechanism in superconducting TiN~\cite{Driessen12}, the interpretation is complicated by a high level of non-magnetic disorder. Thin TiN films, typically fabricated for superconducting devices, are characterized by a relatively small Ioffe-Regel parameter of $k_Fl\lesssim 10$, where $k_F$ is the Fermi wave-vector and $l$ is the carrier mean-free path. Thus, a gradual suppression of the superconductivity in thin films is attributed to the interplay of disorder and interactions~\cite{Finkelstein94,Gantmakher2010,Delacour11,Sacepe2011} or the Berezinskii-Kosterlitz-Thouless phase transition~\cite{Baturina2012}. In order to clarify the role of the magnetic disorder in thin films, one needs to isolate this effect by studying epitaxial films exhibiting excellent electrical properties.  

In this work, we focus on the electronic and superconducting properties of the epitaxial TiN films with an exceptionally low level of non-magnetic disorder, $k_Fl\sim500$. At decreasing film thickness $(d)$ in the range $200$\,nm\,$>d>\,3$\,nm, we observe an almost ten-fold increase of the  residual resistivity, which manifests a predominant contribution of diffusive surface scattering for films thinner than $20$\,nm. At the same time, the superconducting critical temperature in thin films is reduced by over a factor of three when compared to the bulk value in TiN. In contrast to previous experiments, the high structural and thus electrical quality of the materials studied allows us to rule out the possible impact of non-magnetic disorder on the superconductivity. We theoretically confirm that a minute amount of magnetic scattering centers, residing mainly near the surface of the film and that are irrelevant to normal state transport, can account for the suppression of the superconductivity of small thickness films. Our results imply that magnetic defects with a surface density of about $10^{12}\,{\rm cm}^{-2}$ reside within the naturally oxidized top layer of TiN, qualitatively similar to other materials~\cite{Vranken1988,Pierre2002,Anton2013,Kumar2016,Rogachev06, Proslier08}.

\section{\label{methods}Fabrication and Measurement setup}

TiN films were grown on a $\left\langle 111\right\rangle$ c-sapphire substrate at a temperature of $800^{\circ}C$ by DC reactive magnetron sputtering from a $99.999\%$  pure Ti target. The growth was performed in an argon-nitrogen environment at a pressure of $5$\,mTorr and an Ar\,:\,N$_2$ flow ratio  of 2\,:\,8\,sccm. The films with different $d$ are divided chronologically in two sets (1 and 2), each set grown without opening the chamber. Between the two growth processes the chamber was opened and the Ti target replaced. Between the subsequent TiN runs during the deposition period, no other material was deposited.

The electronic properties of the unpatterned TiN films from the two sets were obtained by means of variable-angle spectroscopic ellipsometry~\citep{Shah17} at room temperature (data for plasma frequency $\omega_p$) and also by resistance measurement in a home-made $^4\rm He$ variable temperature insert and a cryo-dilution refrigerator. The resistance measurements are carried out with the 370 AC Lakeshore resistance bridge at a bias current of $316$\,nA and less. 

Sheet resistance $R_s^{300K}$ of films with $d=3,10,100$\,nm is measured by van der Pauw method at room temperature. T-dependences of resistance $R(T)$ and $RRR=R^{300K}/R^{10K}$ are measured in the quasi-four probe configuration. At low T, $R_s^{10K}$ is extracted using the relation $R_s^{10K}=R_s^{300K} RRR^{-1}$. Films with $d=4,5,20,200$\,nm are patterned in Hall-bridges, and $R_s(T)$ with $RRR$ are investigated in four-probe configuration. The uncertainty in the measurement of $R_{s}^{300K}$ is determined from a statistics in different samples of the same thickness.

\section{\label{sec:results}Results and discussion}

The epitaxial TiN films are known to exhibit single-crystalline order~\cite{Naik2012,Kinsey14}. In the following, we start from a demonstration of exceptional metallic properties of our films and investigate the electron-phonon scattering and disorder scattering contributions to the film resistivity. This enables us to evaluate the thickness of the oxide ("dead") layer on the surface of the film and the $d$-dependent mean-free path at low $T$. Next, we study the superconducting properties and analyze the suppression of the superconducting critical temperature $T_\mathrm{c}$ with decreasing film thickness. Using the Abrikosov-Gorkov theory~\cite{AG61} we estimate the density of the magnetic defects in fabricated films and observe that in the thin-film limit, the magnetic disorder has a predominantly surface origin. The electronic properties of the films are summarized in Table~\ref{tab:table1}. 

\begin{figure}
\includegraphics[width = 1\linewidth]{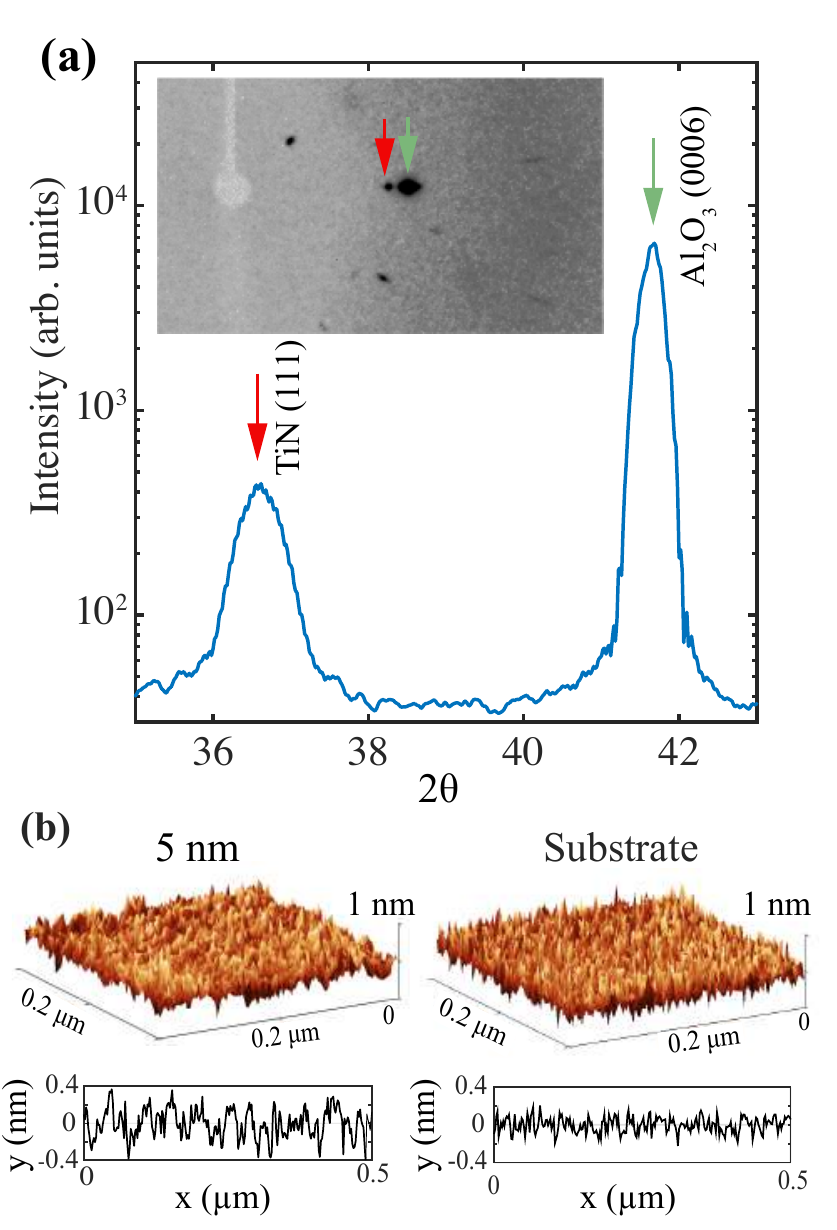} 
\caption{\label{fig:1} Characterization of  the epitaxial TiN films. (a): X-ray diffraction study of a 20\,nm TiN film. Body: coupled $\omega-2\theta$ scan, obtained with a Shimadzu XRD-7000 S diffractometer (Cu source, $\rm K_{\alpha 1}=1.5406\,\AA$). Inset: gray-scale reciprocal space map obtained with an Oxford diffraction Gemini–R diffractometer (Mo source, $\rm K_{\alpha 1}=0.7093\,\AA$) for $\Delta\omega=3^\circ$ around the central value of $\omega=9.05^\circ$. In both plots the vertical arrows mark the positions of the TiN (111) reflex from the film and the (much stronger) $\rm Al_2O_3$ (0006) reflex from the sapphire substrate. (b): AFM images of the 5\,nm TiN film (on the lhs) and the sapphire substrate (on the rhs) along with the corresponding representative cuts. The rms roughness of the TiN film is below $0.25$\,nm, which corresponds to the atomically smooth surface.}
\end{figure}

\begin{figure}
\includegraphics[width = 0.9\linewidth]{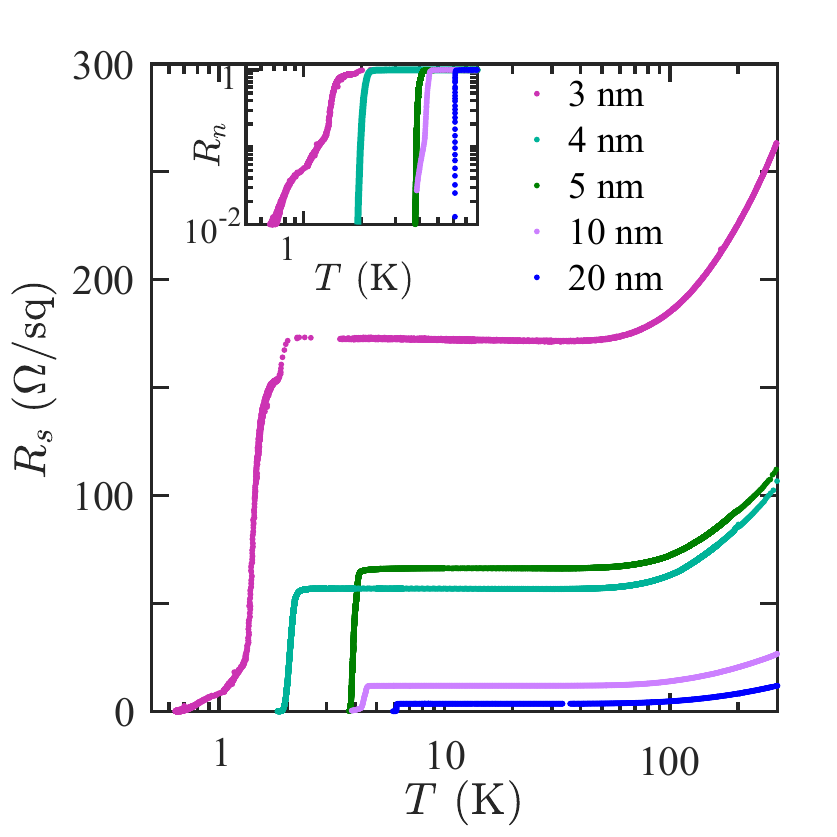} 
\caption{\label{fig:2} The temperature ($T$) dependence of sheet resistance ($R_s$) for five films with different thickness: $20$\,nm (set 1), $10$\,nm (set 1), $5$\,nm (set 1),  $4$\,nm (set 2) and $3$\,nm (set 2). Body: The $R_s(T)$-dependence in the wide $T$ range.  The critical temperatures of transition to superconducting state are $5.6$\,K, $4.5$\,K, $3.8$\,K, $1.9$\,K, $1.4$\,K for $20$\,nm, $10$\,nm, $5$\,nm, $4$\,nm, $3$\,nm, respectively. Inset: The $R(T)$-dependence for the TiN films at low temperatures in log-log scale. The data are obtained with four-probe (quasi) method, the bias current is $316$\,nA.}
\end{figure}

Structural characterization of our TiN films is summarized in Fig.~\ref{fig:1}. In the body of panel Fig.~\ref{fig:1}(a) we plot the X-ray diffraction data for a coupled $\omega-2\theta$ scan of a 20\,nm TiN film. Here we identify and mark by the vertical arrows the two  main reflexes of TiN (111) and $\rm Al_2O_3$ (0006), respectively, for the film and for the substrate. The gray-scale plot in the inset demonstrates that both these reflexes (marked by the same arrows) correspond to localized points in the reciprocal space, evidencing that our epitaxial TiN films are monocrystalline. In Fig.~\ref{fig:1}(b) we plot the atomic force microscope image of a $5$\,nm-thick TiN film and its  substrate along with the representative cuts. The root mean square surface roughness of the TiN film is below $0.25$\,nm, that corresponds to the atomically smooth surface. The details of an additional X-ray photoelectron spectroscopy (XPS) of our samples are given in the Supplemental Material~\citep{Suppl_Data}. The XPS results are obtained on the base of the method  described in Ref.~\cite{Lubenchenko2018}.

Figure~\ref{fig:2} summarizes the electronic transport properties of the fabricated TiN films of different thicknesses. Here, we plot the experimental $T$-dependencies of the sheet resistance $R_{s}$ for TiN films  in zero magnetic field. At decreasing $T$, the $R_{s}$ initially drops linearly and saturates at a residual resistance below about 50\,K. This linear in temperature behavior is fully consistent with the high-$T$ asymptote of the Bloch-Gr\"{u}neisen formula that holds in normal metals down to temperatures $T\propto \theta_D/3$~\cite{Ziman2001}, where $\theta_D$ is the Debye temperature. Our estimate of $\theta_D$ for TiN is in the range of $480 - 600$\,K (see Supplemental Material~\citep{Suppl_Data} for details), in agreement with the previously reported values~\citep{Spengler1978}. This metallic behavior is typical for all studied films and reveals substantial electron-phonon scattering contribution down to a few nanometer film thickness. The residual resistance ratio, listed in Table~\ref{tab:table1}, reaches $RRR\,=\,7$ emphasizing the high quality of films. The room-$T$ resistivity $\rho^{300K}\equiv R_{s}^{300K}d$ attains the values as low as $\rho^{300K}\sim20\,\rm \mu\Omega\cdot$cm for $d\geq100$\,nm, which is similar to the best reported results in thin films~\cite{Chawla2013,Torgovkin2018} as well as in a thick single crystal~\citep{Spengler1978}. This similarity is not surprising given the fact that $\rho^{300K}$ is determined by the phonon scattering, rather than disorder, once again emphasizing the quality of the material and its conceptual difference from the disordered TiN films investigated in most previous works~\citep{Sacepe2008, Baturina2012,Driessen12}. 
 \begin{table}
\caption{\label{tab:table1}Parameters of the TiN films.}
\resizebox{0.45\textwidth}{!}{
 \begin{tabular}{c c c c c c c c c}
\cline{2-9}
\hline \hline 
\\
 & $d$ & $R_{s}^{300K}$ & $RRR$&$T_c$ & $\xi_0$&$\omega_p(300K)$&$\tau_{tr}(10K)$& $l(10K)$\\
& (nm) & (Ohm/sq)  && (K) & (nm) &(eV)& (fs)& (nm) \\ \hline

\multicolumn{1}{ c|}{\multirow{3}{*}{\rotatebox[origin=c]{90}{Set\,1}}}  & 20 & 9.8$\pm$1.0 & 3.2 & 5.6 & 22&7.02& 18 & 8.0 \\ \cline{2-9} 
\multicolumn{1}{ c|}{}                &10 &  26.6$\pm$2.3 &2.2& 4.5 &  22 & 6.85& 11 & 5.5\\ \cline{2-9} 
\multicolumn{1}{c|}{}                & 5 &  103$\pm$6.3 & 1.7 & 3.8&  20&6.44&7 &3.5 \\ \hline
\multicolumn{1}{c|}{\multirow{4}{*}{\rotatebox[origin=c]{90}{Set\,2}}} & 200 & 0.99$\pm$0.2 & 7.0 & 4.5  &  41&6.41& 43 &19 \\ \cline{2-9} 
\multicolumn{1}{c|}{}                & 100 & 1.9$\pm$0.1 & 6.2&  4.5&  42&6.97&33 &17 \\ \cline{2-9} 
\multicolumn{1}{c|}{}                 & 4 &  130$\pm$18.5 &1.9 & 1.9 &  31&7.01&6 &3.5  \\ \cline{2-9} 
\multicolumn{1}{c|}{}                &  3 &  264$\pm$14.5 & 1.5 &1.4 &  26&7.02&4 &2.2  \\ 
\hline \hline 
\end{tabular}}
\end{table}

\begin{figure}
\includegraphics[width = 0.9\linewidth]{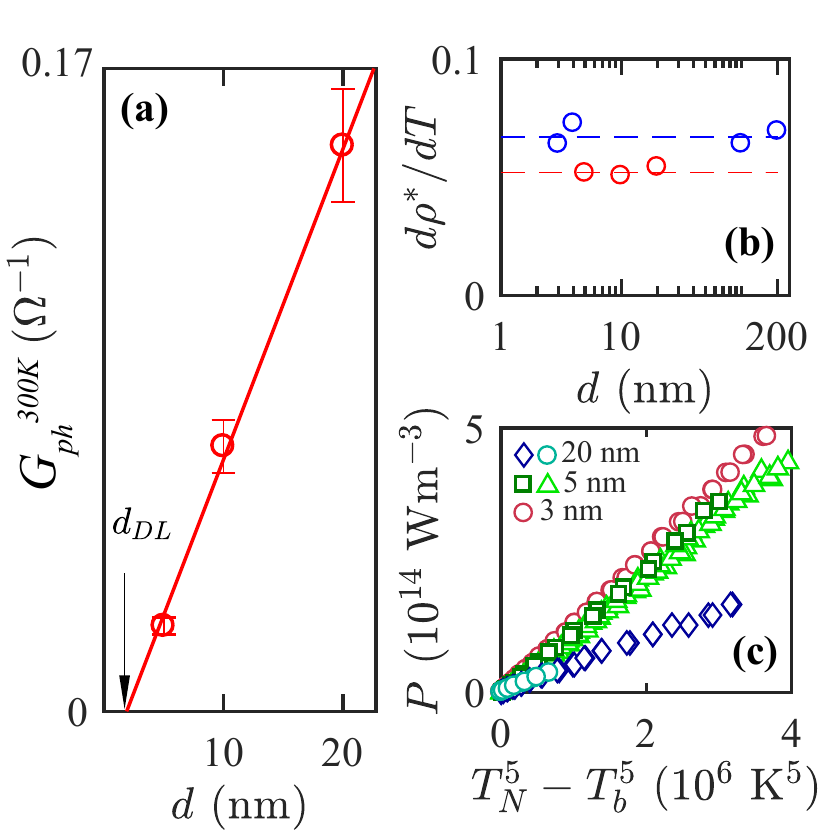}
\caption{\label{fig:3} The analysis of the strength of the electron-phonon (e-ph) coupling with film thickness decrease. (a): the thickness dependence of the electron-phonon contribution to the conductance, $G_{\rm ph}^{300K}=(R_{s}^{300K}-R_s^{10K})^{-1}$. The linear fit of these data provides the  estimate of thickness of an insulating dead-layer at the surface of TiN film: $d_{DL} = 1.9$ nm.  (b): the slope of the high-$T$ linear part of the $T$-dependence of the resistivity as a function of $d$. The resistivity is corrected for the finite dead-layer thickness as $\rho^*(T) = R_{s}(d-d_{DL})$. The data indicates a minor variation of the strength of the electron-phonon scattering in the two sets (red dots are for set 1 and blue dots - for set 2). 
(c): The heat power flow, normalized to volume, vs. $(T_e^5- T_{b}^5)$, where $T_{\rm N}$ the noise temperature, $T_{b}$ the bath temperature. The data are obtained for five samples with thicknesses of $20$\,nm, $5$\,nm and $3$\,nm. }
\end{figure}

We now investigate the electron-phonon (e-ph) interaction  in our films in more detail, which allows us to evaluate the thickness of the dead layer on the surface of the films and understand the $d$-dependence of the e-ph coupling strength. In Figs.~\ref{fig:3}(a), we analyze the phononic contribution to the TiN film conductance at room-$T$, defined as $G_{\rm ph}^{300K}=(R_{s}^{300K}-R_s^{10K})^{-1}$. Plotted as a function of $d$, the $G_{\rm ph}^{300K}$ shows a linear dependence with a finite intercept around $d_{DL}\approx1.9\pm\,0.5$\,nm for the set\,1 (see the guide line).  In other words, the phonon-induced conductance scales linearly with $d^*$, where $d^*=d-d_{DL}$, indicating that a minor size-effect observed in $G_{\rm ph}^{300K}$ vs $d$ is consistent with a trivial decrease of the effective film thickness. Most likely, the insulating dead-layer at the surface of our TiN films consists of a mixture of titanium oxide and oxynitride~\citep{Guler2013, Zgrabik2015}. The presence of such a dead-layer is consistent with the XPS spectra (see Supplemental Material Table~1~\citep{Suppl_Data} for the XPS film profile). As shown in Fig.~\ref{fig:3}(b), where we plot the slope of the high-$T$ linear part of the $T$-dependence of the resistivity as a function of $d$, the correction for the dead layer thickness, introduced as $\rho^*= R_{s}d^*$, is capable to account for the observed $d$-dependence of the (e-ph) scattering in both film sets.
 
Further insight into the e-ph coupling at low temperatures, in the residual resistance range, was obtained via noise thermometry~\cite{Roukes1985}. In this experiment, five devices made of  $3$-nm, $5$-nm and $20$-nm thick films were dc biased and the resulting noise temperature ($T_{\rm N}$) measured with the help of a home-made noise amplification stage (see Fig.~\ref{fig:6} in {\rm Appendix A} for the details). The dependence of the  $T_{\rm N}$ on the Joule power per unit volume, $P$, is demonstrated in Fig.~\ref{fig:3}(c). This dependence is very well described by the heat outflow law $P=\Sigma_{\rm e-ph}(T_{\rm N}^5-T_{b}^5)$, where $T_{b}$ is the bath temperature and $\Sigma_{\rm e-ph}$ is the effective e-ph coupling. The exponent of $5$ in this expression corresponds to the case of e-ph relaxation in clean metals. The measured $\Sigma_{\rm e-ph}$ increases at decreasing $d$ from $20$\,nm to $3$\,nm, roughly by a factor of $2$ or even stronger, if one takes the finite $d_{DL}$ into account. Note that $\Sigma_{\rm e-ph}$ is directly proportional to the coupling strength in the BCS theory of the superconductivity~\citep{Allen1987}. As such, the noise thermometry indicates stronger BCS-coupling in thinner films, which is opposite to the trend observed in $T_c$ as a function of $d$ in the data of Fig.~\ref{fig:2}. This conclusion will be important for our discussion of the superconducting properties below.

\begin{figure}
\includegraphics[width = 0.9\linewidth]{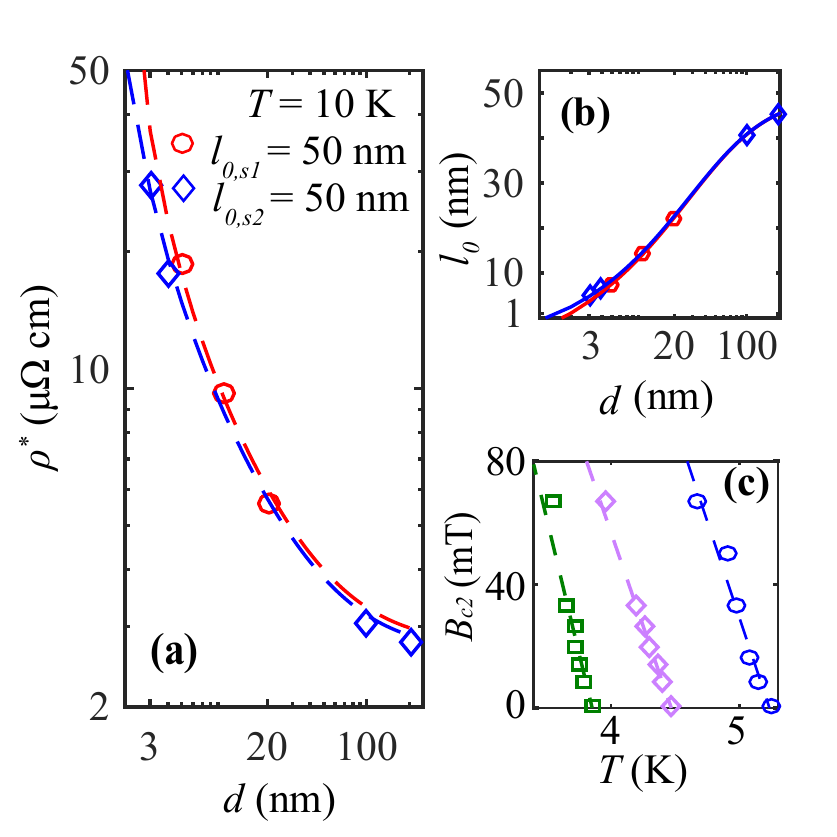}
\caption{\label{fig:4}  Size effect in the residual resistivity. (a): Resistivity as a function of film thickness. The red circles correspond to data for the set\,1, the blue diamonds  - for the set\,2. The values of resistivity are corrected assuming the dead layer as $\rho_0^* = R_{s}^{10K}(d - d_{DL})$. The red and blue dashed lines are fits FS-model (Eq.~\ref{Eq1}) assuming diffusive surface scattering. (b): The disorder limited mean-free path $l=l_0\rho_0^*/\rho^*$ in TiN films (shown by symbols) as function of $d$, along with the FS-based fits for the two sets.  (c): The typical $T$ dependencies of the second critical magnetic field $B_{c2}$ for three films with different thickness: $5$\,nm, $10$\,nm and $20$\,nm (set 1, from left to right). The $B(T)$-dependencies are obtained from the shift of the superconducting transition temperature in a perpendicular  magnetic field identified  as the point where resistance  is a half of the resistance in the normal state. These data are used to determine the Ginzburg-Landau superconducting  coherence length $\xi_0$ at $T\,=\,0\,K$.}
\end{figure}

 Unlike the case of e-ph conductance $G_{\rm ph}^{300K}$, the analysis of residual resistivity reveals a much stronger size-effect in dependence of the film thickness.  At low $T$, the mean-free path increases and we observe the size effect at decreasing $d$, with a transition from the dominant bulk scattering in thick films to the surface scattering in thin films.  Fig.~\ref{fig:4} shows that in both sets the residual resistivity measured at $T=10$\,K increases  at least by a factor of four for decreasing $d$.  Assuming diffusive surface scattering, we fit the data using the Fuchs-Sondheimer model (FS-model)~\citep{Fuchs1938, Sondheimer1952}:

\begin{multline}
\frac{\rho_0^*}{\rho^*} =1 - \frac{3l_0}{2d^*}\int_1^{\infty} \left( \frac{1}{t^3} - \frac{1}{t^5} \right) (1 - e^{- t d^*/l_0})dt, 
\label{Eq1}
\end{multline}
 where $d^*=d-d_{DL}$ and the fit parameters $l_0$  and  $\rho_0^*$ are, respectively, the mean-free path and the resistivity in the thick film limit.  The best fits shown by the dashed lines in Fig.~\ref{fig:4}(a) correspond to $l_0 = 50\pm 5$\,nm and  $\rho_0^*= 2.7\pm 0.2$\,$\rm\mu\Omega\cdot cm$ for the set\,1  and $l_0 = 50\pm 5$\,nm and  $\rho_0^*= 2.6\pm 0.2$\,$\rm\mu\Omega\cdot cm$ for the set\,2. This procedure allows us to evaluate the $d$-dependent mean-free path $l=l_0\rho_0^*/\rho^*$ in our films, shown by symbols in Fig.~\ref{fig:4}(b), and estimate the Ioffe-Regel parameter as high as  $k_Fl_0=\sqrt{3\pi^2\hbar l_0/e^2\rho_0^*}\approx 500$ in the thick film limit. This estimate is two times bigger than the mean-free paths obtained  independently from the data on transport relaxation time and diffusion coefficient extracted from the measured  $\rho^*$, plasma frequency $\omega_p$ and the Ginzburg-Landau superconducting  coherence length (see Table~\ref{tab:table1}). The plasma frequency is measured by ellipsometry at room temperature, the coherence length is determined using the relation $(\xi_0)^2=-\Phi_0(dB_{c2}/dT)^{-1}/2\pi T_c$ from  the temperature dependencies of the second critical magnetic field $B_{c2} (T)$, see the Figure~\ref{fig:4}(c). The transport relaxation time is estimated as $\tau_{tr}  =1/(\rho^{10K}\omega_p^2\varepsilon_0)$, assuming that $\omega_p$ is temperature independent~\citep{Vertchenko2019}. Note that in our analysis of the size-effect in the residual resistance we excluded the Mayadas-Shatzkes (MS) model~\citep{Mayadas1969}, which focuses on the scattering of electrons at grain boundaries in polycrystalline thin films. The negligible granularity in our epitaxial films directly follows from our XRD and topography data in Fig.~\ref{fig:1}. Consistently, when applied to our data, the MS model returns negligible contribution of scattering at grain boundaries (see Supplemental Material Fig. 3~\citep{Suppl_Data} for the details).

We conclude the transport studies in the normal state of our TiN films by evaluating the charge carrier density, $n$, from the product  $\rho^*l$. In spirit of Ref.~\cite{Chawla2013}, we use the FS model fits of Fig.~\ref{fig:4}(a) and the free-electron expression~\cite{Sondheimer1952} $\rho\,l=(3\pi^2)^{1/3}n^{-2/3}\hbar/e^2$, where $\hbar$ is the Planck constant and $e$ is the elementary charge. In this way we obtain $n\approx 2.9\times10^{22}$\,cm$^{-3}$, which is the same order of magnitude compared to the density $n = 8\times10^{22}$\,cm$^{-3}$ expected for a single electron per
Ti atom, as well as to the experimental value of $n\approx 5\times10^{22} {\rm cm}^{-3}$ obtained from the Hall effect measurements in nominally identical films~\cite{Shah17}. This observation is also consistent with the fact that the Bloch-Gr\"{u}neisen  temperature is close to the Debye temperature in our analysis of the e-ph scattering (see Supplemental Material Fig.S4~\citep{Suppl_Data}), that excludes a diluted metal scenario~\cite{Hwang2019}. Altogether, our analysis does not support the conclusions of Ref.~\cite{Chawla2013} that the charge transport in epitaxial TiN films is dominated by the minority carriers from slightly filled bands. 

Next, we analyze the superconducting properties of the fabricated TiN films. We observe a sharp transition to the superconducting state that occurs below a few Kelvin at $T=T_c$, see the inset in Fig.~\ref{fig:2}. For simplicity, we determine the critical temperature $T_c$ in Table~\ref{tab:table1} as the point where the resistance halves compared to $R_s^{10K}$. Note that the variation of the $T_c$ in the experiment by far exceeds the width of the resistive transition, therefore the  results discussed below are insensitive to a criterion used to define the transition point~\citep{Varlamov18}. In both sets, the $T_c$ values considerably diminish as the $d$ is reduced. The critical temperature is systematically lower within the set\,2 and  varies by more than a factor of $3$ for the thinnest films (see Table~\ref{tab:table1}). Note that the effect of decreasing $T_c$ occurs in high quality films with $R_{s}\ll \hbar/e^2$, that is the films are far away from superconductor-insulator transition~\citep{Gantmakher2010}. In this case the non-magnetic disorder does not affect $T_c$.  Thus, we also exclude the Berezinskii-Kosterlitz-Thouless phase transition~\cite{Baturina2012} and the impact of Coulomb interactions~\citep{Finkelstein87}, responsible for a decrease of $T_c$ in thin dirty superconducting films (see Fig.~\ref{fig:7} in {\rm Appendix B} for details). We also eliminate possible effects of the reduced carrier density and/or BCS-coupling strength in thin films~\citep{Bourgeois03, Hsu91}, because the observed trends in $\omega_p$ (Table~\ref{tab:table1}) and the e-ph coupling (Figs.~\ref{fig:3}(b) and~\ref{fig:3}(c)) are absent or opposite to that for $T_c$.

\begin{figure}
\includegraphics[width = 0.9\linewidth]{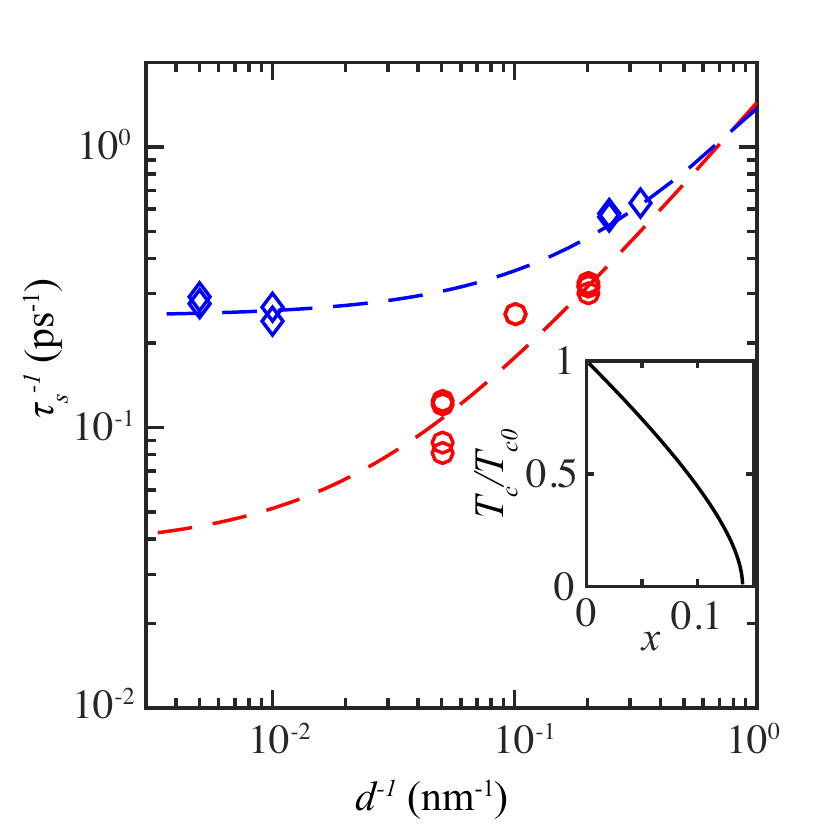}
\caption{\label{fig:5} The illustration of a detrimental effect of magnetic disorder on $T_c$ in thin films. Body: the estimated spin-flip scattering rate $\tau_s^{-1}$ as a function of an inverse thickness $1/d$. Inset: the solid line demonstrates the dependence of the normalized $T_c$ as a function of normalized spin-flip rate $x=\hbar/(2\pi k_BT_{c}^0\tau_s)$ given by Eq.~(\ref{Eq2}). This dependence is used to convert the measured $T_c$ for each studied TiN film to the spin-flip rate. The fitting of the data ($\tau_s^{-1}$  vs. $d^{-1}$) demonstrates that the dominant contribution to the spin-flip scattering in thin films originates from the near surface magnetic disorder: $N_{M}\sim a/(v_F\tau_s) = N_b + N_s\times a/d$, where $a\approx 0.4$\,nm~\citep{Allmaier09} is the TiN lattice constant, $v_F$ is the Fermi velocity, $N_b$ and $N_s$ are the effective bulk and the surface number of the magnetic scatterers, respectively.  The dashed lines  demonstrate the best fits for the two sets, obtained with the bulk defect number of $N_b=3\times 10^{-5}$ for the set\,1 and $N_b=2.5\times 10^{-4}$ for the set\,2 and the surface defect number of $N_s=2.8\times 10^{-3}$, the same for both sets.}
\end{figure}

Both the observed differences in $T_c$ between the two sets and its decrease upon the reduction of $d$ can be explained by the presence of a minute amount of magnetic disorder, that has a well-known detrimental effect on $T_c$ owing to pair breaking spin-flip scattering~\citep{AG61}. The spin-flip scattering time $\tau_s$ and the critical temperature of the superconducting transition $T_c$ are related via  the Abrikosov-Gorkov (AG) equation~\citep{Ludwig71, AG61}:

\begin{equation}
\ln\left(\frac{T_{c}^0}{T_c}\right) = \psi\left( \frac{1}{2}+\frac{\hbar}{2\pi k_BT_{c}^0\tau_s}\right) - \psi\left(\frac{1}{2}\right), 
\label{Eq2}
\end{equation}
 where $\psi(x)$ is the digamma function, and $T_{c}^0$ is the critical temperature in the absence of magnetic disorder. The solid line in the inset of Fig.~\ref{fig:5} demonstrates the dependence of the normalized $T_c$ as a function of normalized spin-flip rate $x=\hbar/(2\pi k_BT_{c}^0\tau_s)$ given by Eq.~(\ref{Eq2}). This dependence is used to extract the spin-flip rate from the measured $T_c$ for each TiN film studied. 
 
For both sets of samples we have assumed the same $T_{c}^0=6$\,K, that is the highest reported value of the critical temperature in TiN~\cite{Spengler1978}. Fig.~\ref{fig:5}(body) presents the dependence of the spin-flip scattering rate  $\tau_s^{-1}$ as a function of an inverse thickness $d^{-1}$. Note the time-scale of $\tau_s$ falls in the range of $1-15$\,ps, that is roughly three orders of magnitude longer compared to the transport scattering time $\tau_{tr}$  for scattering off the non-magnetic disorder (Table~\ref{tab:table1}). Therefore, we once again exclude the role of the non-magnetic disorder in the scaling of transition temperatures in thin films. 

Fig.~\ref{fig:5} demonstrates that the spin-flip scattering rate increases at decreasing $d$. We argue that this is consistent with the surface magnetic disorder that dominates in thin films. For $d<\xi_0$, the superconductivity is sensitive to the total volume density of the magnetic scatterers regardless of their distribution within the cross-section of the film. Hence, the $d$-dependence of the $\tau_s^{-1}$ indicates that extra spin-flip scattering in thin films originates from the magnetic disorder residing near the surface. 

 We apply a simplified theoretical model capable  to qualitatively reconcile our data. Using the data of  Fig.~\ref{fig:5}, we  extract the effective density of the magnetic scatterers $N_{M}$, including the bulk $N_b$ and the surface $N_s$ contributions explicitly. It is convenient to normalize the numbers $N_{M},\,N_b$ and $N_s$, respectively, per 3D and 2D unit cells in TiN, such that $N_{M} = N_b + N_s\times a/d$, where $a\approx 0.4$\,nm~\citep{Allmaier09} is the TiN lattice constant. The relation between the $N_{M}$ and the spin-flip scattering rate reads $N_{M}\sim a/(v_F\tau_s)$, where  $v_F$ is the Fermi velocity. The dashed lines in Fig.~\ref{fig:5} demonstrate the best fits for the two sets, obtained with $N_b=3\times 10^{-5}$ for the set\,1 and $N_b=2.5\times 10^{-4}$ for the set\,2 and $N_s=2.8\times 10^{-3}$, the same for both sets. These estimates are obtained using the average value of the Fermi velocity $v_F \approx 4\times10^7$\,cm/s extracted from experimental values of electron diffusivity $D$ and the electron scattering time $\tau_{tr}$ as $v_F\sim \sqrt{D/\tau_{tr}}$ (see Supplemental Material Fig.~S6~\cite{Suppl_Data}). While different values of $N_b$ can account for a growth related variation between the sets, the same value of $N_s$ indicates that the observed drop of $T_c$ at decreasing $d$  is an important systematic effect in thin epitaxial TiN films. The values of $N_s$ provide us with an estimate of the surface density of magnetic defects that is as small as $a^{-2}N_s=10^{12}\,{\rm cm}^{-2}$, at least an order of magnitude smaller in comparison with a typical density of the surface magnetic moments ($\sim 5\times 10^{13}$ cm$^{-2}$), reported for Al, Nb and NbN superconductors~\citep{Koch07, Sendelbach09, Proslier2011, kumar16, deGraaf2018}. Note, however, that relevant for the $T_c$ reduction are only those magnetic scatterers that strongly couple to the conduction electrons. This could, at least partly, explain the obtained very small density of the surface magnetic disorder in our analysis.

Finally, we discuss possible  microscopic origin of the surface magnetic disorder. It should be noted that magnetic materials were never used in the TiN growth chamber, thereby a trivial contamination with paramagnetic impurities is excluded in the studied films. The surface character of the magnetic scattering in thin films indicates the importance of the TiN interfaces either with the substrate on the bottom or with the dead-layer on the top. Similar to the observations in copper~\cite{Vranken1988,Pierre2003}, aluminum~\cite{Kumar2016}  and niobium~\citep{Proslier08} films, we propose that the naturally oxidized top layer can be responsible for the magnetic disorder in our films. The magnetic moments in this case can originate from the unpaired $3d$ electrons bound to $\rm Ti^{+3}\text{-}O_V$ defect complexes~\cite{Zhou09}, where $\rm O_V$ is the oxygen vacancy, which can result even in a  room-$T$ ferromagnetism in  TiO$_2$~\citep{Hong06,Yoon2006,Drera_PhD,Drera10}. Recently, a long range magnetic ordering in non-stoichiometric epitaxial TiN$_{1-x}$ with $x=0.12\pm0.02$ was revealed in Ref.~\cite{Gupta2019}, which originates from the RKKY interaction between the unpaired localized spins mediated by nitrogen vacancies. Such spins, yet in much smaller concentration, can also be considered as possible magnetic scatterers in our TiN films, both in bulk and on the surface. In the end, it is worth mentioning that in our analysis of the $T_c$ reduction at decreasing $d$ we have ignored peculiarities of the band structure in TiN, which is argued to be a correlated material close to the Mott-insulator phase transition point~\citep{Allmaier09}. Possible interplay between the band structure and magnetic disorder in thin TiN films is an intriguing target for future experiments.

 \section{\label{conclusion}Conclusions}
 
  In summary, we analyzed the electronic properties of the epitaxial TiN films of exceptional quality ($k_Fl\sim500$), which exhibit a size effect in resistivity and the reduction of the superconducting critical temperature with decreasing film thickness. High structural and electronic quality of the films allows us to relate the latter effect to the presence of a minute concentration ($\rm\sim10^{12}\,cm^{-2}$) of magnetic scatterers within the $\sim2$\,nm thick dead-layer on the top of TiN films. The observed surface magnetic disorder can be related to the oxygen vacancies in naturally oxidized TiN films, representing a fundamental limiting factor for their performance in the superconducting state.

\begin{acknowledgements}
We acknowledge valuable discussions with P.I.~Arseev, M.V.~Feigelman, T.M.~Klapwijk, D.V.~Shovkun and M.A.~Skvortsov. We are grateful to S.V.~Simonov and S.L. Shestakov for their assistance with the X-ray studies, and N.S. Kaurova for her assistance with the AFM measurements. The authors also acknowledge N. Dilley for preliminary measurements of superconducting critical temperatures and critical fields in the TiN films. The Purdue team acknowledges support from the U.S. Department of Energy, Office of Basic Energy Sciences, Division of Materials Sciences and Engineering under Award DE SC0017717 (growth of TiN films and measurement of plasma frequency). The transport and noise measurements were funded by the Russian Science Foundation project No.17-72-30036. The reciprocal state map of Fig.1(a) was obtained within the state task of the ISSP RAS. The surface analysis (AFM and XPS) was funded by RFBR project number 16-29-11779. The theoretical analysis was supported by the Grant of the President RF No.~MK-1308.2019.2.
\end{acknowledgements}

\section*{\label{appendix a} APPENDIX A: STUDY OF ELECTRON-PHONON HEAT TRANSFER IN EPITAXIAL TIN FILMS}

The noise thermometry is used to study the heat transfer between the electron system and the heat bath in the normal state of superconducting materials~\citep{Roukes1985}. In such measurements, the sample is biased with a DC current that causes to Joule heating of the electronic system, and thereby noise increases. The noise temperature $T_{\rm N}$, obtained from the Johnson-Nyquist relation, $S_{\rm I} = 4k_B T_{\rm N}/R$, is considered as the electron temperature $T_e$, and the phonon temperature $T_{\rm ph}$ is taken as the bath temperature $T_{\rm b}$.

\begin{figure}[h!]
\includegraphics[width = 0.9\linewidth]{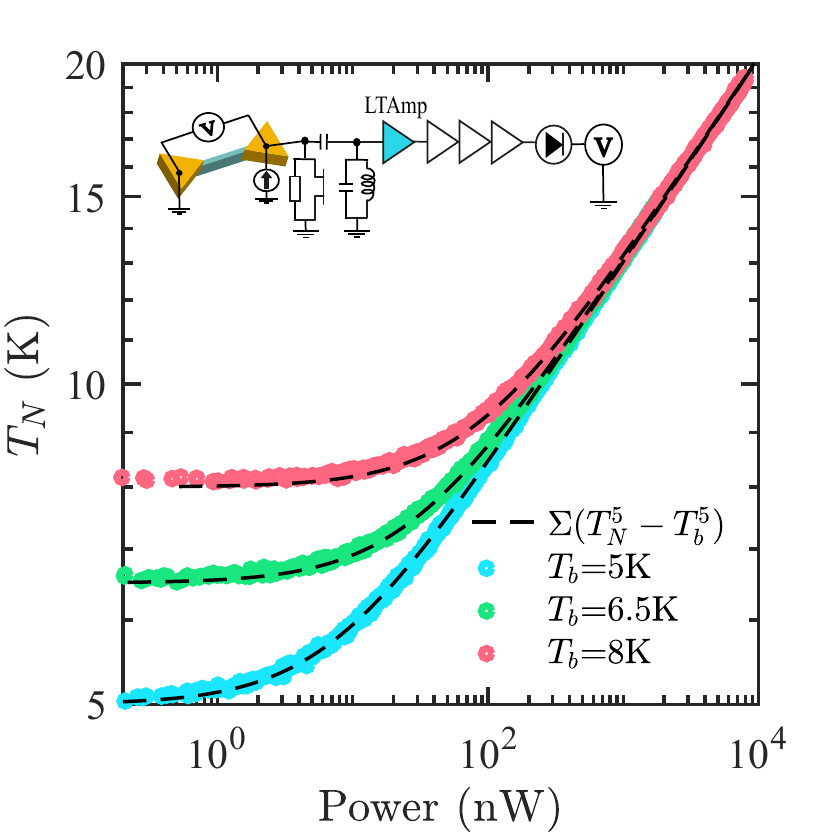}
\caption{\label{fig:6} Study of electron-phonon heat transfer in epitaxial TiN films. Body:\,The noise temperature $T_N$ vs. Joule power $P_{DC}$. The data are presented for samples with thicknesses of $5$\,nm\,. Insert:\,The experimental setup for noise thermometry.}
\end{figure}

For samples with length $L>l_{\rm e-ph}$, where $l_{\rm e-ph}$ is electron-phonon length, the heat flow out rate can provide information about the electron-phonon interaction. The data in Fig.~\ref{fig:6} are presented for sample with thickness of $5$\,nm. For the samples the experimental data follows the heat flow out law, $P_{\rm DC}\,=\,\Sigma_{\rm e-ph}\,V(T^n_{\rm N} - T^n_{\rm b})$, where $P_{\rm DC}$ is the Joule power dissipated in the sample, $V$ is the volume of the sample, $\Sigma_{\rm e-ph}$ is the electron-phonon coupling constant. We observed that the exponent in heat flow law is $n\simeq5$ for all samples, which is typical for the case of pure metals and in the absence of a phonon bottleneck effect.

An experimental setup for noise thermometry is presented in the insert of Figure~\ref{fig:6}. The setup, built inside a closed cycle refrigerator Bluefors LD-400, consists of a RF resonant-tank circuit (with a resonance frequency of 10\,MHz) including a high-impedance low-noise amplifier at the 4\,K-stage (with gain $>6$dB and noise $S_{\rm amp}\sim10^{-26}$\,A$^2$/Hz), a cascade of low-noise amplifiers at $300$\,K, an active bandpass filter and a power detector.

\section*{\label{appendix b} APPENDIX B: DISORDER EFFECTS ON CRITICAL TEMPERATURE}

\begin{figure}[h!]
\includegraphics[width = 0.9\linewidth]{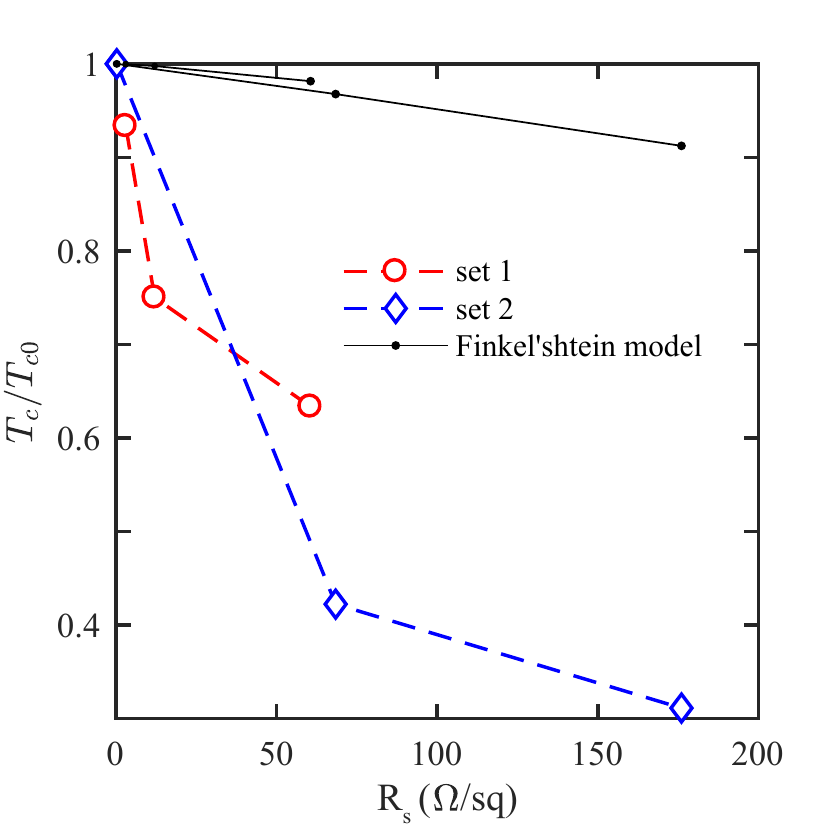} 
\caption{ \label{fig:7} Disorder effects on critical temperature. Plot of the ratio $T_c/T_{c0}$ as a function of the film resistance per square $R_{s}$. Data $R_{s}$ is determined above the superconducting transition (at $10$\,K). The data are compared with the Finkel'stein model for case of  weak disorder in homogeneous superconducting films. The used model parameters are $\gamma\approx 6.7$ and $\tau = l^2/(N D)\approx 60$\,fs.}
\end{figure}

We compare our results with predictions of a weak disorder model in homogeneous superconducting films established by Finkel'stein~\citep{Finkelstein87} (see Figure.~\ref{fig:7}). The suppression of superconductivity is driven by impurities that reinforce Coulomb and spin interactions. The critical temperature $T_c$ is expressed as a function of sheet resistance $R_{s}$ and the elastic diffusion time $\tau$:
\begin{equation}
\frac{T_c}{T_{c0}} = e^\gamma\left(\frac{1/\gamma - \sqrt{t/2}+t/4}{1/\gamma +\sqrt{t/2}+t/4}\right)^{1/\sqrt{2t}},
\label{Eq3}
\end{equation}
with $t =R_{s}e^2/(\pi h)$, $\gamma = ln(h/(k_BT_{c0}\tau))$ and $T_{c0} = 6 K$ (for set 1) and $T_{c0}=4.5 K$ (for set 2).

 \bibliographystyle{apsrev4-1}
\nocite{apsrev41Control} 
\bibliography{titanium_nitride2017}

\end{document}